\documentclass[12pt,a4paper]{article}
\usepackage{a4wide}
\usepackage{amsmath}
\usepackage{amssymb}
\usepackage{epsfig}
\usepackage{subfigure}
\usepackage{exscale}
\usepackage{float}
\usepackage{bbm}
\usepackage[numbers,sort&compress]{natbib}
\usepackage{pst-plot, pstricks,pst-math}
\usepackage{fancybox,amssymb,color}
\usepackage{graphicx}
\usepackage{pstricks, color, graphicx, epsfig, psfrag}
\usepackage{amsfonts,amsmath,amssymb,slashed}
\usepackage{dsfont}
\usepackage{bbm,bm}
\usepackage{fancyhdr, a4wide}
\usepackage[english]{babel}
\usepackage{subfigure}
\makeatletter
\@addtoreset{equation}{section}
\makeatother

\title{Pion Pressure in a Magnetic Field}

\author{Christoph P.\ Hofmann$^a$ \\ \\
\normalsize{$^a$ Facultad de Ciencias, Universidad de Colima} \\
\vspace{0.3cm}
\normalsize {Bernal D\'iaz del Castillo 340, Colima C.P.\ 28045, Mexico} \\}

\begin{document}

\maketitle

\begin{abstract} \normalsize

While the partition function for QCD in a magnetic field $H$ has been calculated before within chiral perturbation theory up to two-loop
order, our investigation relies on an alternative representation for the Bose functions which allows for a clear-cut expansion of
thermodynamic quantities in the chiral limit. We first focus on the pion-pion interaction in the pressure and show that -- depending on
magnetic field strength, temperature and pion mass -- it may be attractive or repulsive. We then analyze the thermodynamic properties in
the chiral limit and provide explicit two-loop representations for the pressure in the weak magnetic field limit $|qH| \ll T^2$.

\end{abstract}

\maketitle

\section{Introduction}
\label{Intro}

The low-energy properties of quantum chromodynamics (QCD) can be understood on the basis of its relevant low-energy degrees of freedom: the
Goldstone bosons. This is the path pursued by chiral perturbation theory (CHPT) and indeed, the low-temperature properties of QCD in a
magnetic field have been explored within CHPT in many studies up to two-loop order \citep{SS97,Aga00,Aga01,AS01,CMW07,Aga08,AF08,Aga09,
And12a,And12b,BK17,AS18b}. Other approaches to finite temperature QCD in magnetic fields include lattice QCD \citep{EMS10,EN11,BBEFKKSS12a,
BBKKP12,BBEFKKSS12b,BBCCEKPS12c,BBEGS13,BBCKS14,BBEKS14,BEMNS14,IMPS14,EMNS18,EGKKP19}, Nambu-Jona-Lasinio model-based studies \citep{GR11,
AA13,FCMPS14,FCLFP14,FCP14,FIPQ14,RLP16,ZFL16}, and other techniques \citep{CW09,MCF10,NK11,FR11,FP12,BEK13,End13,OS13,BFP13,RTG13,BBES14,
CFS14,OS14a,OS14c,HPS14,MP15,OS15,SO15,KK15,PEP16,TDES16,BHM17,RP17,AS18a,Bra18,RP18,KGBHM19}. Yet more references can be found in the
review Ref.~\citep{ANT16}.

Recently, the present author has pointed out that the low-temperature expansion of the quark condensate in a weak magnetic field and
in the chiral limit has not been  properly derived, and has provided the correct series in Ref.~\citep{Hof19}. The analysis was based on an
alternative representation for the Bose functions which was the key to derive the proper series in a transparent manner. Relying on this
coordinate space representation, here we take the analysis up to the two-loop level. This does not merely correspond to rederiving or
rephrasing known results for QCD in magnetic fields in an alternative framework. Rather, our analysis goes beyond the literature by (i)
analyzing how the nature of the pion-pion interaction in the pressure -- repulsive or attractive -- is affected by the magnetic field, as
well as temperature and pion mass, and (ii) by providing the low-temperature series for the pressure in weak magnetic fields
($|qH| \ll T^2$) in the chiral limit.

In terms of the dressed pions at zero temperature, the low-temperature expansion of the pressure in a magnetic field takes a remarkably
simple form. Non-interacting pions yield the well-known $T^4$-contribution, while interaction effects enter at order $T^6$. In the chiral
limit ($M\to0$) -- irrespective of whether or not the magnetic field is present -- the two-loop interaction contribution vanishes. In the
case $M\neq0$, the pion-pion interaction in the pressure is mostly attractive, but may become repulsive at low temperatures as the magnetic
field strength grows. In general, sign and magnitude of the interaction depend on the actual values of temperature, magnetic field, and
pion masses in a nontrivial way, as we illustrate in various figures.

In the chiral limit, the low-temperature expansion of the pressure in a weak magnetic field $H$ is dominated by terms involving
$\epsilon^{3/2}, \epsilon^2 \log\epsilon$ and $\epsilon^2$, where $\epsilon = |q H|/T^2$ is the relevant expansion parameter and $q$ is the
electric charge of the pion.

The article is organized as follows. The evaluation of the QCD partition function in a magnetic field up to two-loop order within chiral
perturbation theory is presented in Sec.~\ref{CHPT}. The nature of the pion-pion interaction in the pressure -- attractive or repulsive --
is analyzed in Sec.~\ref{pressureNature}. We then focus in Sec.~\ref{chiralLimit} on the thermodynamic properties in the chiral limit and
provide the weak magnetic field expansion of the pressure to arbitrary order in $\epsilon = |q H|/T^2$. Finally, in Sec.~\ref{conclusions}
we conclude. While details on the two-loop CHPT evaluation are discussed in Appendix \ref{appendixA}, the rather technical analysis of the
pressure in weak magnetic fields in the chiral limit is presented in Appendix \ref{appendixB}.

\section{Chiral Perturbation Theory Evaluation}
\label{CHPT}

\subsection{Preliminaries}
\label{preliminaries}

Surveys of chiral perturbation theory have been provided on many occasions (see, e.g., Refs.~\citep{Leu95,Sch03}) -- in what follows we
only touch upon the very basic elements to set the notation. Throughout the study, we refer to the isospin limit $m_u = m_d$.

The effective pion fields $\pi^i(x)$ appear in the SU(2) matrix $U(x)$,
\begin{equation}
U(x) =\exp(i \tau^i \pi^i(x)/F) \, , \qquad i=1,2,3 \, ,
\end{equation}
where $\tau^i$ are Pauli matrices and $F$ represents the pion decay constant at tree level. The leading piece in the effective Lagrangian
is of momentum order $p^2$ and takes the form
\begin{equation}
\label{L2}
{\cal L}^2_{eff} = \mbox{$ \frac{1}{4}$} F^2 Tr \Big[ {(D_{\mu} U)}^\dagger (D_{\mu} U) - M^2 (U + U^\dagger) \Big] \, .
\end{equation}
Here $M$ is the pion mass at tree level . The magnetic field $H$ enters via the covariant derivative
\begin{equation}
D_{\mu} U = \partial_\mu U + i [Q,U] A^{EM}_\mu \, ,
\end{equation}
where $Q$ is the charge matrix of the quarks, $Q=diag(2/3,-1/3)e$. The gauge field $A^{EM}_\mu=(0,0,-H x_1,0)$ contains the (constant)
magnetic field \citep{ANT16}.

\begin{figure}
\begin{center}
\includegraphics[width=15cm]{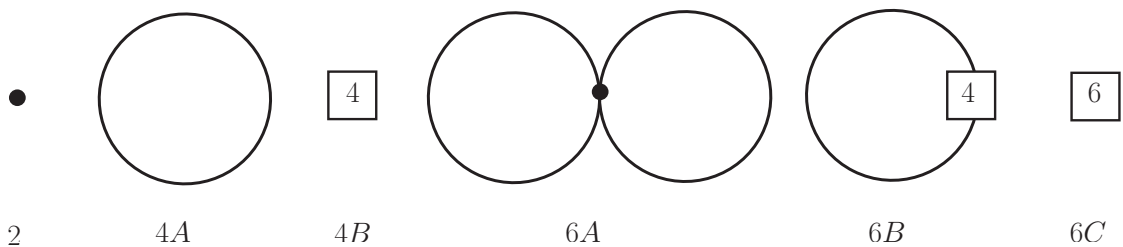}
\end{center}
\caption{QCD partition function diagrams up to order $p^6 \propto T^6$. The filled circle stands for ${\cal L}^2_{eff}$, the numbers $4$ and
$6$ in the boxes represent ${\cal L}^4_{eff}$ and ${\cal L}^6_{eff}$.}
\label{figure1}
\end{figure}

In the present analysis, we also need higher-order pieces of the effective Lagrangian, namely ${\cal L}^4_{eff}$ and ${\cal L}^6_{eff}$. The
explicit structure is given, e.g., in Refs.~\citep{GL84,BCE00}. The relevant Feynman diagrams for the partition function up to two-loop
order $p^6$ are shown in Fig.~\ref{figure1}. The lines represent thermal propagators which either correspond to the charged pions or the
neutral pion. In fact, the dimensionally regularized zero-temperature propagator $\Delta^0(x)$ for the neutral pion in Euclidean space
takes the familiar form
\begin{equation}
\label{regprop}
\Delta^0(x) = (2 \pi)^{-d} \int {\mbox{d}}^d p \, e^{ipx} (M^2 + p^2)^{-1}
= {\int}_{\!\!\! 0}^{\infty} \mbox{d} \rho \, (4 \pi \rho)^{-d/2} e^{- \rho M^2 - x^2/{4 \rho}} \, .
\end{equation}
On the other hand, the dimensionally regularized zero-temperature propagator $\Delta^{\pm}(x)$ for the charged pions does involve the
magnetic field. In Euclidean space, as derived in Ref.~\citep{Hof19}, it amounts to
\begin{equation}
\Delta^{\pm}(x) = \frac{|qH|}{{(4 \pi)}^{\frac{d}{2}}} \, e^{-s_{\perp} |qH| x_1 x_2/2} \, {\int}_{\!\!\! 0}^{\infty} \mbox{d} \rho \,
\frac{\rho^{-\frac{d}{2}+1} e^{-\rho M^2}}{\sinh(|qH| \rho)} \,
\exp\Bigg[\!-\frac{x^2_4 + x^2_3}{4\rho} - \frac{|qH| (x^2_1 + x^2_2)}{4 \tanh(|qH|\rho)}\Bigg] \, ,
\end{equation}
where
\begin{equation}
s_{\perp} = sign(qH) \, .
\end{equation}
In either case -- for neutral and charged pions -- the thermal propagators are obtained by the summing over zero-temperature propagators as
\begin{equation}
\label{ThermalPropagator}
G(x) = \sum_{n = - \infty}^{\infty} \Delta({\vec x}, x_4 + n \beta) \, , \qquad \beta = \frac{1}{T} \, .
\end{equation}
In the evaluation of the partition function diagrams displayed in Fig.~\ref{figure1}, as we will see, thermal propagators only have to be
considered at the origin $x$=0. It is furthermore advantageous to isolate the zero-temperature pieces $\Delta^{\pm}$ and $\Delta^0$ in the
thermal propagators via
\begin{eqnarray}
\label{decomposition}
& & G^{\pm}(0) \equiv G^{\pm}_1 = \Delta^{\pm}(0) + g^{\pm}_1(M,T,H) \, , \nonumber \\
& & G^0(0) \equiv G^0_1 = \Delta^0(0) + g_1(M,T,0) \, .
\end{eqnarray}
The quantities $g^{\pm}_1(M,T,H)$ and $g_1(M,T,0)$ are kinematical functions that describe the purely finite-temperature part. They are
embedded in the more general class of Bose functions defined by
\begin{eqnarray}
\label{BoseMTH}
g^{\pm}_r(M,T,H) & = & \frac{T^{d-2r-2}}{{(4 \pi)}^{r+1}} \, |qH| {\int}_{\!\!\! 0}^{\infty} \mbox{d} \rho
\frac{\rho^{r-\frac{d}{2}}}{\sinh ( |qH| \rho/ 4 \pi T^2 )}
\, \exp\Big( -\frac{M^2}{4 \pi T^2} \rho \Big) \Bigg[ S \Big( \frac{1}{\rho} \Big) -1 \Bigg] \, , \nonumber \\
g_r(M,T,0) & = & \frac{T^{d-2r}}{{(4 \pi)}^r} \, {\int}_{\!\!\! 0}^{\infty}  \mbox{d} \rho \rho^{r-\frac{d}{2}-1} \, \exp\Big( -\frac{M^2}{4 \pi T^2}
\rho \Big) \Bigg[ S\Big( \frac{1}{\rho} \Big) -1 \Bigg] \, , \nonumber \\
& & S(z) = \sum_{n=-\infty}^{\infty} \exp(- \pi n^2 z) \, ,
\end{eqnarray}
where $S(z)$ is the Jacobi theta function. Note that $g_r(M,T,0)$ does not involve the magnetic field. In order to facilitate the
subsequent analysis, in the Bose functions $g^{\pm}_r(M,T,H)$ for the charged pions, we extract the $H$=0 part as
\begin{equation}
g^{\pm}_r(M,T,H) = {\tilde g}_r(M,T,H) + g_r(M,T,0) \, ,
\end{equation}
where solely
\begin{eqnarray}
\label{BoseZeroH}
{\tilde g}_r(M,T,H)& = & \frac{T^{d-2r-2}}{{(4 \pi)}^{r+1}} \, |qH| {\int}_{\!\!\! 0}^{\infty} \mbox{d} \rho \rho^{r-\frac{d}{2}} \,
\Bigg( \frac{1}{\sinh(|qH| \rho /4 \pi T^2)} - \frac{4 \pi T^2}{|qH| \rho} \Bigg) \nonumber \\
& & \times \, \exp\Big( -\frac{M^2}{4 \pi T^2} \rho \Big) \Bigg[ S\Big( \frac{1}{\rho} \Big) -1 \Bigg]
\end{eqnarray}
contains the magnetic field. These two types of kinematical functions -- ${\tilde g}_r(M,T,H)$ and $g_r(M,T,0)$ -- constitute the basic
building blocks in our analysis. The decomposition of the thermal propagators into $T$=0 and finite-$T$ pieces then results in
\begin{eqnarray}
\label{decomp2}
& & G^{\pm}_1 = \Delta^{\pm}(0) + {\tilde g}_1(M,T,H) + g_1(M,T,0) \, , \nonumber \\
& & G^0_1 = \Delta^0(0) + g_1(M,T,0) \, .
\end{eqnarray}

As a low-energy effective field theory, chiral perturbation theory describes QCD in the regime where quark masses as small, magnetic fields
are weak and temperatures are low: the quantities $M, H$ and $T$ ought to be small compared to the QCD scale
$\Lambda_{QCD} \approx 1 \, GeV$. While ratios between these parameters in principle can have any value, in the present analysis, the limits
$M/T \to 0$ (chiral limit at fixed temperature) and $|qH| \ll T^2$ (weak magnetic field limit) are of particular interest.

\subsection{Free Energy Density up to Order $p^6$}
\label{fed}

The one-loop free energy density (order $p^4$) -- in coordinate space representation -- has been derived in Ref.~\citep{Hof19}. The final
renormalized expression reads
\begin{equation}
\label{freeEDp4}
z_{2+4A+4B} = z^{[4]}_0 - \mbox{$ \frac{3}{2}$} g_0(M,T,0) - {\tilde g}_0(M,T,H) \, .
\end{equation}
The zero-temperature part $z^{[4]}_0$ is\footnote{The third term in the first parenthesis should read $-\frac{3}{2}$. In Ref.~\citep{Hof19},
Eq.~(A7), it was inadvertently cited as -1.}
\begin{eqnarray}
\label{freeEDp4ZeroT}
z^{[4]}_0 & = & - F^2 M^2 + \frac{M^4}{64 \pi^2} \, \Big({\overline l_3} - 4{\overline h_1} - \frac{3}{2}\Big) + \frac{{|qH|}^2}{96 \pi^2} \,
( {\overline h_2} - 1) \\
& & - \frac{{|qH|}^2}{16 \pi^2} {\int}_{\!\!\! 0}^{\infty} \mbox{d} \rho \rho^{-2} \Big( \frac{1}{\sinh(\rho)} - \frac{1}{\rho} + \frac{\rho}{6}
\Big) \, \exp\!\Big( -\frac{M^2}{|qH|} \rho \Big) \, .\nonumber
\end{eqnarray}
Modulo factors of $\gamma_3/32 \pi^2$, $\delta_1/32 \pi^2$, and $\delta_2/32 \pi^2$, the quantities $\overline l_3$, $\overline h_1$, and
$\overline h_2$ represent the running effective coupling constants evaluated at the renormalization scale $\mu = M_{\pi}$
($M_{\pi} \approx 140 \, MeV$) -- details can be found, e.g., in Ref.~\citep{GL89}.

At the two-loop level (order $p^6$) the three partition function diagrams $6A\!-\!C$ have to be evaluated -- this is done in appendix
\ref{appendixA}. In terms of the tree-level pion mass $M$, the outcome is
\begin{eqnarray}
\label{freeEDp6}
z_{6A+6B+6C} & = & z^{[6]}_0 +\frac{3 M^2}{8 F^2} \, {(g_1)}^2 + \frac{M^2}{2 F^2} \, g_1 {\tilde g}_1 \nonumber \\
& & + g_1 \Bigg[ -\frac{3 {\overline l}_3}{64 \pi^2} \, \frac{M^4}{F^2} + \, \frac{M^2}{2 F^2} \, K_1
+ \frac{{\overline l}_6 - {\overline l}_5}{48 \pi^2} \frac{{|qH|}^2}{F^2} \Bigg] \nonumber \\
& & + {\tilde g}_1 \Bigg[ -\frac{{\overline l}_3}{32 \pi^2} \, \frac{M^4}{F^2}
+ \frac{{\overline l}_6 - {\overline l}_5}{48 \pi^2} \frac{{|qH|}^2}{F^2} \Bigg] \, ,
\end{eqnarray}
where the integral $K_1$ is defined in Eq.~(\ref{defK1}). Since we are interested in the behavior of the system at finite temperature, the
explicit structure of the $T$=0 contribution $z^{[6]}_0$ is not needed here.

Let us have a closer look at the terms linear in $g_1$ and ${\tilde g}_1$. First notice that the kinematical functions $g_1$ and
${\tilde g}_1$ are related to $g_0$ and ${\tilde g}_0$ through
\begin{equation}
g_1 = - \frac{\partial g_0}{\partial M^2} \, , \qquad {\tilde g}_1 = - \frac{\partial {\tilde g}_0}{\partial M^2} \, .
\end{equation}
In presence of a magnetic field, the mass of a charged pion ($M^{\pm}_{\pi}$) is different from the mass of a neutral pion ($M^0_{\pi}$). In
order to determine these masses, we express the kinematical functions $g_0$ and ${\tilde g}_0$ in terms of $M^{\pm}_{\pi}$ and $M^0_{\pi}$
-- instead of $M$. Since only the charged pions are tied to ${\tilde g}_r$\footnote{See Eq.~(\ref{decomp2}).}, we can write
\begin{equation}
\label{taylor1}
{\tilde g}_0(M^{\pm}_{\pi},T,H) = {\tilde g}_0(M,T,H) - {\tilde g}_1(M,T,H) \, {\tilde \epsilon}_1 \, ,
\end{equation}
where ${\tilde \epsilon}_1$ measures the mass square difference
\begin{equation}
{\tilde \epsilon}_1 = {(M^{\pm}_{\pi})}^2 - M^2 \, .
\end{equation}
Comparing with the third line of Eq.~(\ref{freeEDp6}), we identify ${\tilde \epsilon}_1$ as
\begin{equation}
{\tilde \epsilon}_1 = \frac{{\overline l}_6 - {\overline l}_5}{48 \pi^2} \, \frac{{|qH|}^2}{F^2} - \frac{{\overline l}_3}{32 \pi^2} \,
\frac{M^4}{F^2} \, .
\end{equation}
As for $g_r$ -- where all three pions are involved according to Eq.~(\ref{decomp2}) -- we must distinguish between the respective pieces:
for the charged pions we write
\begin{equation}
\label{taylor2}
g_0(M^{\pm}_{\pi},T,0) = g_0(M,T,0) - g_1(M,T,0) \, {\tilde \epsilon}_1 \, ,
\end{equation}
while for the neutral pion we have
\begin{equation}
\label{taylor3}
g_0(M^0_{\pi},T,0) = g_0(M,T,0) - g_1(M,T,0) \, {\epsilon}_1 \, .
\end{equation}
The quantity ${\epsilon}_1$ measures the mass square difference
\begin{equation}
{\epsilon}_1 = {(M^0_{\pi})}^2 - M^2 \, ,
\end{equation}
and can be identified as
\begin{equation}
{\epsilon}_1 = - \frac{{\overline l}_3}{32 \pi^2} \, \frac{M^4}{F^2} +  \frac{M^2}{F^2} \, K_1 \, .
\end{equation}
As a result, we can read off how the pion masses are affected by the magnetic field,
\begin{eqnarray}
\label{pionMasses}
{(M^{\pm}_{\pi})}^2 & = & M^2_{\pi} + \frac{{\overline l}_6 - {\overline l}_5}{48 \pi^2} \, \frac{{|qH|}^2}{F^2} \, ,
\nonumber \\
{(M^0_{\pi})}^2 & = & M^2_{\pi} + \frac{M^2}{F^2} \, K_1 \, .
\end{eqnarray}
Note that $M_{\pi}$ is the pion mass in zero magnetic field given by
\begin{equation}
\label{Mpi}
M^2_{\pi} = M^2 - \frac{{\overline l}_3}{32 \pi^2} \, \frac{M^4}{F^2} + {\cal O}(M^6) \, .
\end{equation}

The mass relations (\ref{pionMasses}) indeed coincide with those obtained by Andersen in Ref.~\citep{And12b} -- see Eqs.~(3.8)-(3.10) -- in
the zero-temperature limit. It should be pointed out that we consider the pion masses at zero temperature, while in Ref.~\citep{And12b}
finite temperature effects are included as well. As it turns out, to have a clear definition of interaction effects in the thermodynamic
quantities, we must consider the pion masses at $T$=0, i.e., dress the pions at zero temperature according to Eq.~(\ref{pionMasses}).

The result for the total two-loop free energy density simplifies considerably if we now express the kinematical functions $g_0$ and
${\tilde g}_0$ in the one-loop contribution -- Eq.~(\ref{freeEDp4}) -- by the masses $M^{\pm}_{\pi}$ and $M^0_{\pi}$ rather than by $M$.
Using Eqs.~(\ref{taylor1}), (\ref{taylor2}), and (\ref{taylor3}), we obtain
\begin{eqnarray}
\label{fedPhysicalM}
z_{tot} & = & z_0 - g_0(M^{\pm}_{\pi},T,0) -\mbox{$ \frac{1}{2}$} g_0(M^0_{\pi},T,0)- {\tilde g}_0(M^{\pm}_{\pi},T,H) \nonumber \\
& & + \frac{M^2_{\pi}}{2 F^2} \, g_1(M^{\pm}_{\pi},T,0) \, g_1(M^0_{\pi},T,0)
- \frac{M^2_{\pi}}{8 F^2} \, {\Big\{ g_1(M^0_{\pi},T,0)  \Big\}}^2 \nonumber \\
& & + \frac{M^2_{\pi}}{2 F^2} \, g_1(M^0_{\pi},T,0) \, {\tilde g}_1(M^{\pm}_{\pi},T,H) + {\cal O}(p^8) \, ,
\end{eqnarray}
where $z_0$ is the zero-temperature piece. The crucial point is that all terms linear in $g_1(M,T,0)$ and ${\tilde g}_1(M,T,H)$ have been
absorbed into mass renormalization: $M^2 \to {(M^{\pm}_{\pi})}^2, {(M^0_{\pi})}^2$. In particular, the effect of the pion-pion interaction at
finite temperature is solely contained in the terms quadratic in the kinematical functions. It should be noted that the difference between
$M^2_{\pi}$, Eq.~(\ref{Mpi}), and the tree-level mass $M^2$ -- at the order we are considering -- is irrelevant in the coefficients
accompanying the terms quadratic in the kinematical functions, such that it is legitimate write $M^2_{\pi}$.

While the evaluation of the two-loop free energy density in Refs.~\citep{And12a,And12b} is based on a momentum-space representation for the
kinematical functions, here we have used an alternative representation based on coordinate space. The advantage is that the latter approach
allows for a clear-cut expansion of thermodynamic quantities in the chiral limit as we demonstrate below.

\section{Pressure: Nature of Pion-Pion Interaction}
\label{pressureNature}

We now explore the manifestation of the pion-pion interaction in the pressure which we derive from the two-loop free energy density as
\begin{equation}
P = z_0 - z_{tot} \, .
\end{equation}
To make temperature powers in the pressure manifest, we replace the Bose functions $g_r$ and ${\tilde g}_r$ by the dimensionless
kinematical functions $h_r$ and ${\tilde h}_r$ according to
\begin{equation}
h_0 = \frac{g_0}{T^4} \, , \quad  {\tilde h}_0 = \frac{{\tilde g}_0}{T^4} \, , \qquad
h_1 = \frac{g_1}{T^2} \, , \quad  {\tilde h}_1 = \frac{{\tilde g}_1}{T^2} \, ,
\end{equation}
and obtain the low-temperature expansion of the pressure as
\begin{equation}
\label{pressure}
P = p_1(t,m,m_H) \, T^4 + p_2(t,m,m_H) \, T^6  + {\cal O}(T^8) \, ,
\end{equation}
with coefficients
\begin{eqnarray}
\label{coeffp1p2}
p_1(t,m,m_H) & = & h_0(M^{\pm}_{\pi},T,0) + \mbox{$ \frac{1}{2}$} h_0(M^0_{\pi},T,0) + {\tilde h}_0(M^{\pm}_{\pi},T,H) \nonumber \\
p_2(t,m,m_H) & = & - \frac{m^2}{2 t^2 F^2} \, h_1(M^{\pm}_{\pi},T,0) \, h_1(M^0_{\pi},T,0)
+ \frac{m^2}{8 t^2 F^2} \, {\Big\{ h_1(M^0_{\pi},T,0) \Big\}}^2 \nonumber \\
& & - \frac{m^2}{2 t^2 F^2} \, h_1(M^0_{\pi},T,0) \, {\tilde h}_1(M^{\pm}_{\pi},T,H) \, .
\end{eqnarray}
The dimensionless parameters $t, m$, and $m_H$,
\begin{equation}
t = \frac{T}{4 \pi F} \, , \qquad m = \frac{M_{\pi}}{4 \pi F} \, , \qquad m_H = \frac{\sqrt{|q H|}}{4 \pi F}\, , 
\end{equation}
measure temperature, pion mass $M_{\pi}$, Eq.(\ref{Mpi}), and magnetic field strength with respect to the scale
$4 \pi F \approx \Lambda_{QCD}$, i.e.,  with respect to the renormalization group invariant scale $\Lambda_{QCD} \approx 1 GeV$. In the
domain where chiral perturbation theory operates, these parameters are small: more concretely, in the plots below, we will restrict
ourselves to the parameter region $t, m, m_H \lessapprox 0.3$. For the pion masses we use $M_{\pi} = 140 \, MeV$ and, following
Ref.~\citep{FLAG20}, for the pion decay constant we get $F = 85.6 \, MeV$. Finally, according to Ref.~\citep{GL84}, for the combination of
NLO low-energy constants as it appears in the charged pion masses, we take ${\overline l_6} -{\overline l_5} = 2.64$.

The $T^4$-contribution in the low-temperature series for the pressure corresponds to the non-interacting pion gas, while the
pion-pion interaction emerges at order $T^6$. Recall that the Bose functions $h_0$ and $h_1$ do not involve the magnetic field: the effect
of the magnetic field is embedded in the Bose functions ${\tilde h}_0$ and ${\tilde h}_1$. In the chiral limit ($M\to0$), the coefficient
$p_2(t,m,m_H)$ tends to zero: the pion-pion interaction only starts manifesting itself at the three-loop level, as is well-known
for the case $H=0$ (see, e.g., Ref.~\citep{GL89}). However, for $M \neq 0$, the interaction term is present and -- depending on the actual
values of the parameters $t, m$ and $m_H$ -- the pion-pion interaction in the pressure may result attractive or repulsive, as we now
illustrate.

\begin{figure}
\begin{center}
\hbox{
\includegraphics[width=8.0cm]{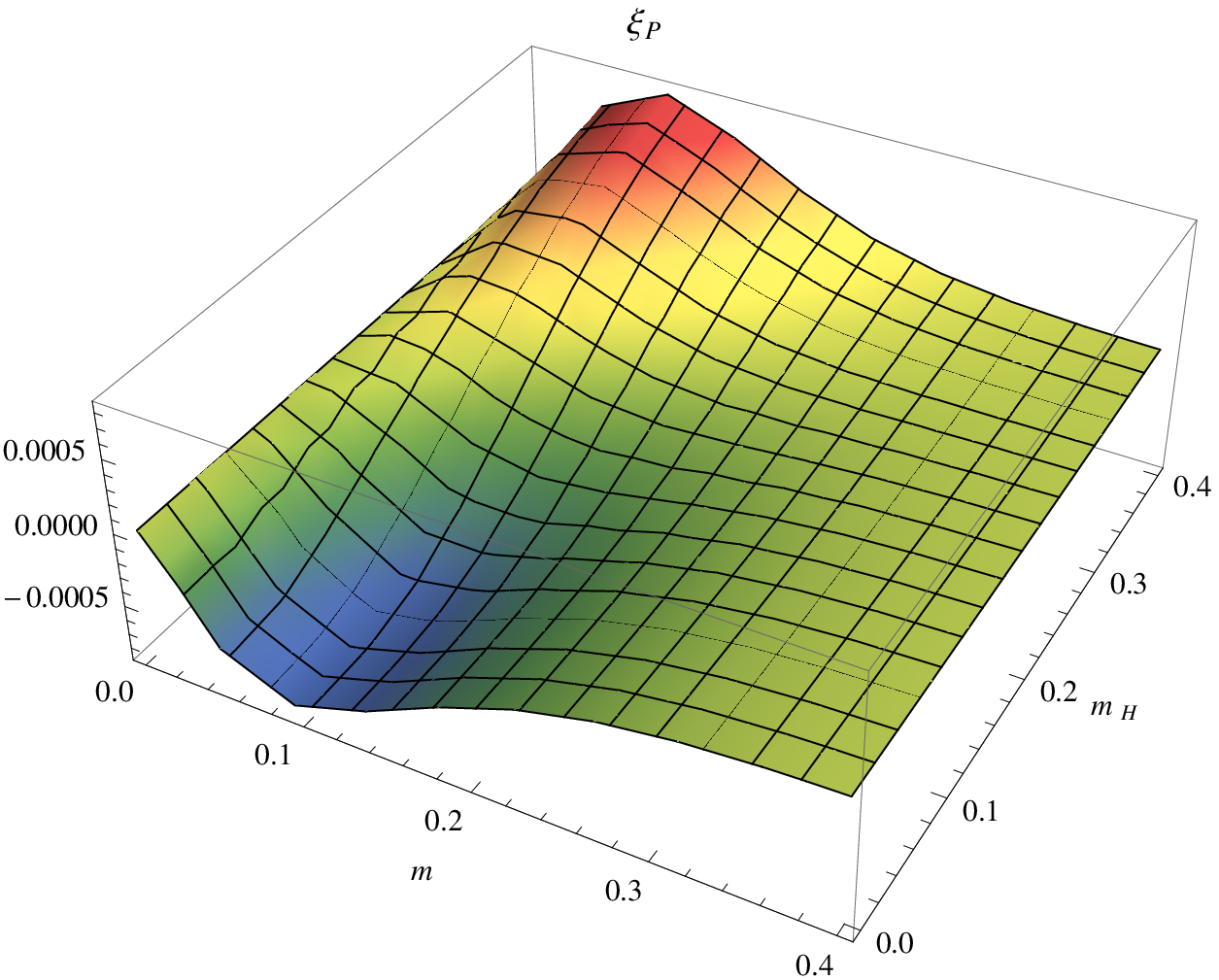} 
\includegraphics[width=8.0cm]{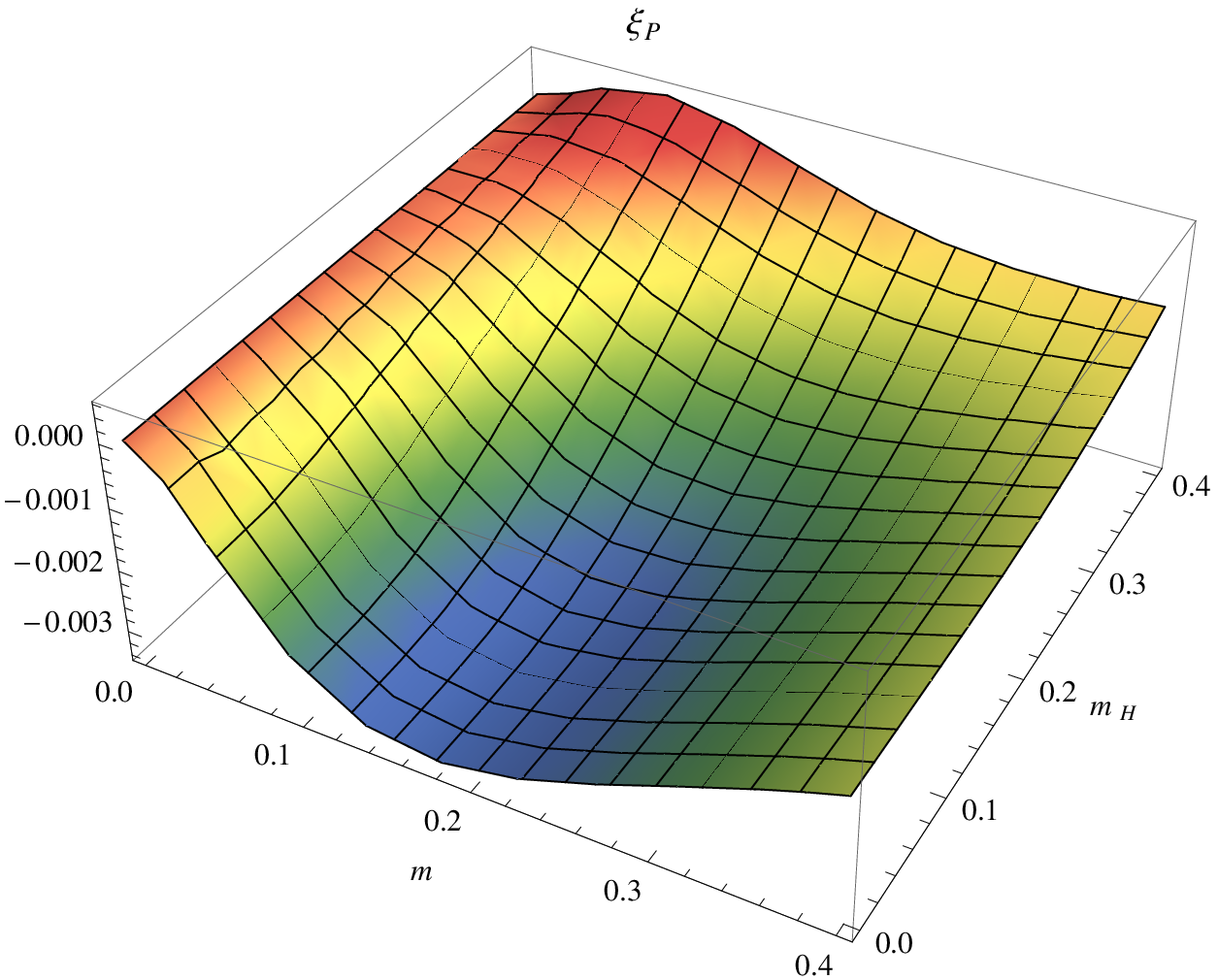}}
\vspace{7mm}
\hbox{
\includegraphics[width=8.0cm]{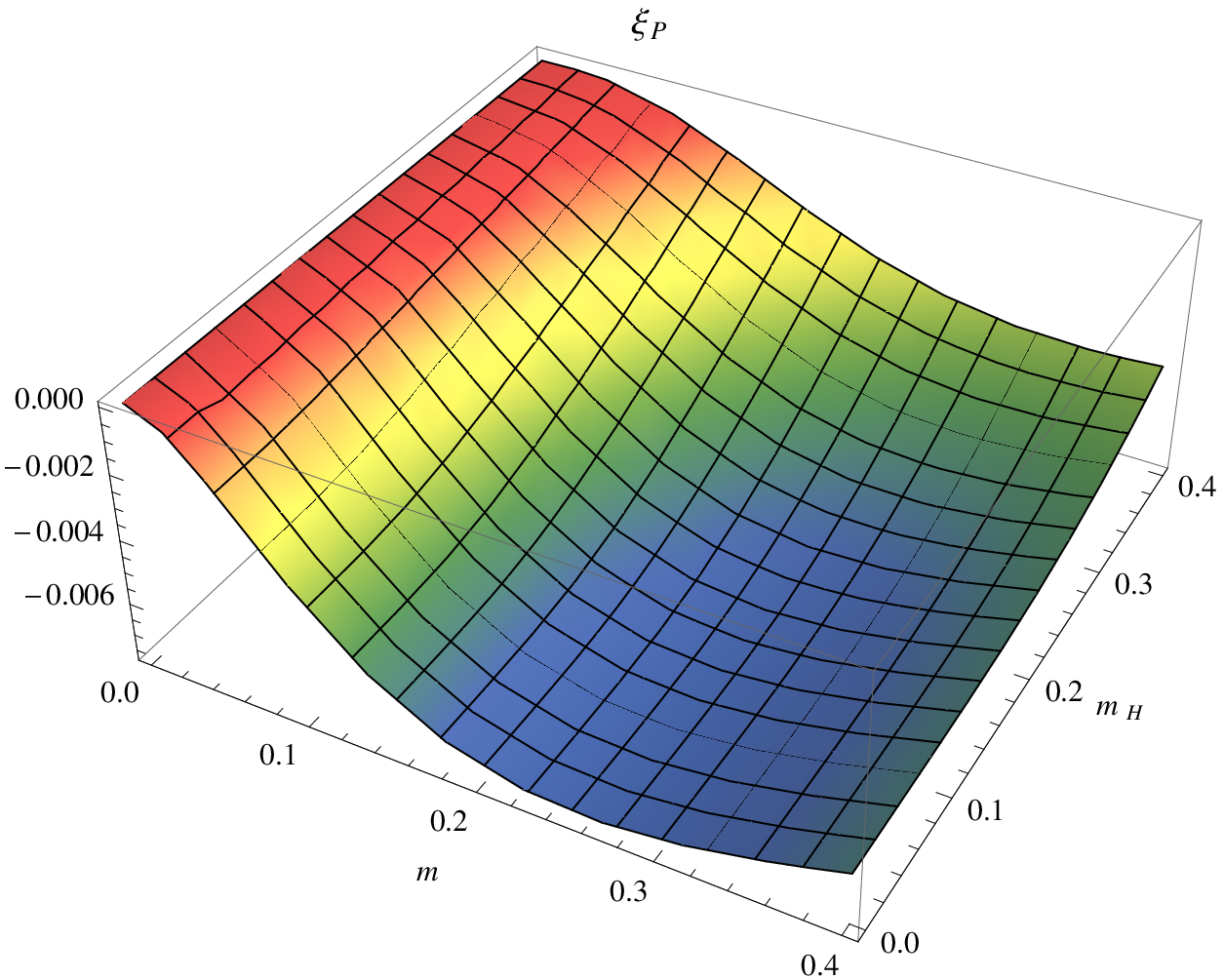}
\includegraphics[width=8.0cm]{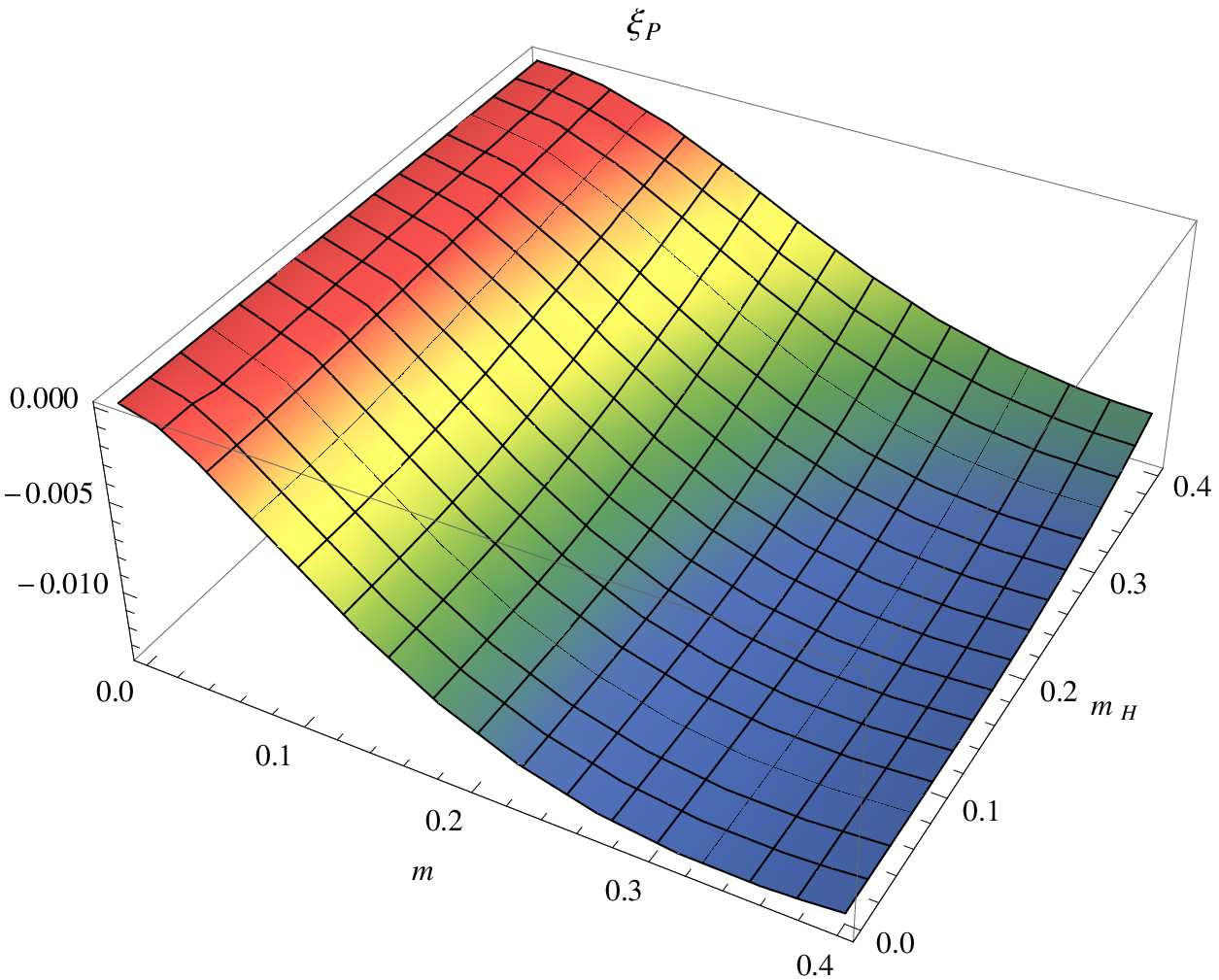}}
\end{center}
\caption{[Color online] Magnitude and sign of the pion-pion interaction in the pressure measured by $\xi_P(t,m,m_H)$ --
Eq.~(\ref{ratioPressure}) -- for the temperatures $T= 53.8 \, MeV, 108 \, MeV$ (upper panel) and  $T= 161 \, MeV,  215 \, MeV$ (lower
panel).}
\label{figure2}
\end{figure}

To get a more quantitative picture, let us consider the dimensionless ratio
\begin{equation}
\label{ratioPressure}
\xi_P(t,m,m_H) = \frac{p_2(t,m,m_H) \, T^2}{p_1(t,m,m_H)}
\end{equation}
that measures magnitude and sign of the pion-pion interaction relative to the non-interacting pion gas contribution. In Fig.~\ref{figure2}
we depict this ratio for the four temperatures $t= \{ 0.05, 0.1, 0.15, 0.2 \} $, or equivalently,
$T= \{ 53.8, 108, 161, 215 \} \, MeV$.

In the limit $M\to0$, irrespective of absence or presence of the magnetic field, the two-loop interaction contribution vanishes. In the
other limit $H\to0$, the interaction in the pressure always is attractive, irrespective of the actual values of the (nonzero) pion masses
and temperature. When the magnetic field is switched on, the attractive pion-pion interaction becomes weaker, but only at low temperatures
and stronger magnetic fields does the pion-pion interaction become repulsive. Overall, the interaction in the pressure is quite small, at
most around one percent compared to the leading free Bose gas contribution. 

\begin{figure}
\begin{center}
\includegraphics[width=12.0cm]{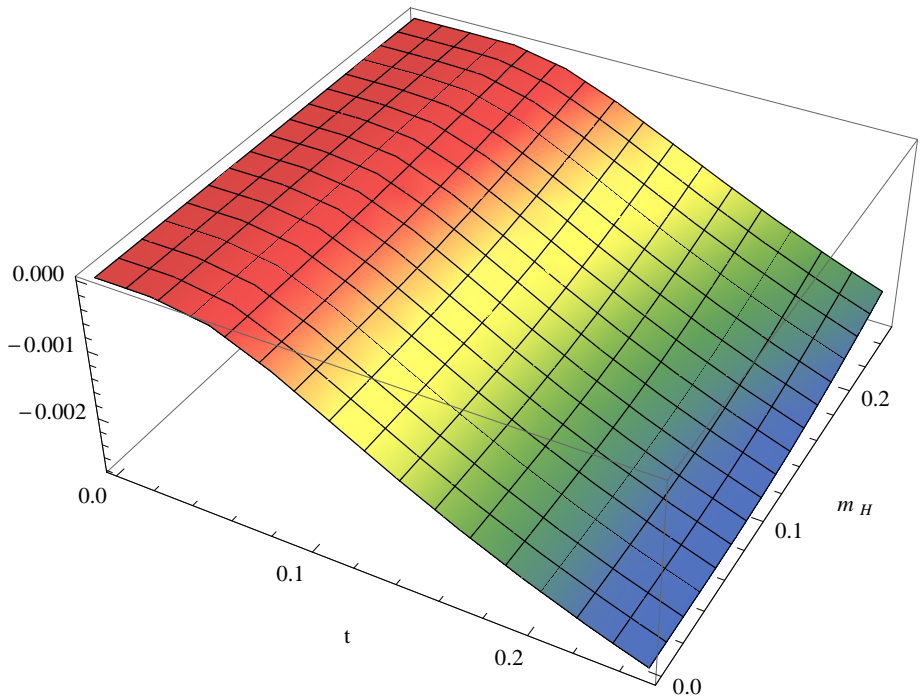}
\end{center}
\caption{[Color online] Magnitude and sign of the pion-pion interaction in the QCD pressure as a function of temperature ($t$) and
magnetic field strength ($m_H$) -- measured by $p_2(t,m,m_H) \, T^2$ -- at the physical value $M_{\pi} = 140 \, MeV$ of the pion masses.}
\label{figure3}
\end{figure}

The case of interest corresponding to the physical value of the pion masses -- $M_{\pi} = 140 \, MeV $, i.e., $m = 0.130$\footnote{Note
that we refer to the isospin limit where all three pions have the same mass (in the absence of the magnetic field).} -- is depicted in
Fig.~\ref{figure3} where we plot the dimensionless two-loop contribution $p_2(t,m,m_H) \, T^2$ as a function of temperature and magnetic
field strength. As the figure suggests, the interaction is purely attractive in the parameter domain ${t,m_H} \le 0.25 $. As the strength of
the magnetic field grows, the attractive interaction gradually becomes weaker. Note that the maximal values for the parameters $t$ and
$m_H$ correspond to $T \approx 269 \, MeV$ and $\sqrt{|q H |} \approx 269 \, MeV$, respectively. In other words, we are already in a region
where temperature and magnetic field strength are no longer small compared to the underlying scale ${\Lambda}_{QCD}$ and the low-temperature
expansion starts to break down.

\section{Pressure  in Weak Magnetic Fields in the Chiral Limit}
\label{chiralLimit}

The objective of Ref.~\citep{Hof19} was to provide the correct low-temperature series for the one-loop quark condensate in weak magnetic
fields in the chiral limit. The corresponding analysis involved the kinematical function ${\tilde g}_1$. Regarding the pressure, the
one-loop contribution involves the kinematical function ${\tilde g}_0$. In appendix \ref{appendixB}, we derive the expansion of this
function in weak magnetic fields in the chiral limit, following analogous strategies as for ${\tilde g}_1$. Based on these results, we now
discuss the structure of the weak magnetic field expansion of the pressure in the chiral limit up to two-loop order which is new to the
best of our knowledge.

In the chiral limit, as stated previously, the pion-pion interaction in the pressure starts showing up only at three-loop order which is
beyond the scope of our investigation. The low-temperature series for the pressure in the chiral limit is hence fixed by the Bose
contribution of order $T^4$ that contains the kinematical function ${\tilde g}_0$. With the weak magnetic field expansion for
${\tilde g}_0$, Eq.~(\ref{seriesTildeg0}), the low-temperature series for the pressure in weak magnetic fields and in the chiral limit
takes the form
\begin{eqnarray}
P & = & \frac{\pi^2}{30} \, T^4 
+ \Bigg\{ -\frac{|I_{\frac{3}{2}}|}{8 \pi^{3/2}} \, \epsilon^{\frac{3}{2}}
- \frac{1}{96 \pi^2} \, \epsilon^2 \, \log\epsilon
+ b_1 \epsilon^2 + {\cal O}(\epsilon^4) \Bigg\} \, T^4 \nonumber \\
& & - ({\overline l_6} - {\overline l_5}) \Bigg\{ \frac{t^2}{144 \pi} \, \epsilon^2
- \frac{t^2 |I_{\frac{1}{2}}|}{96 \pi^{5/2}} \, \epsilon^{\frac{5}{2}}
+ \frac{t^2 \log 2}{192 \pi^3} \, \epsilon^3 + {\cal O}(\epsilon^4) \Bigg\} \, T^4 \nonumber \\
& & + {\cal O}(T^8 \log T) \, ,
\end{eqnarray}
where the relevant expansion parameter $\epsilon \ll 1$ is
\begin{equation}
\epsilon = \frac{|qH|}{T^2} \, .
\end{equation}
The quantities
\begin{equation}
I_{\frac{3}{2}} \approx -0.610499 \, , \qquad  b_1 \approx 0.00581159 \, , \qquad  I_{\frac{1}{2}} \approx -1.516256 \, ,
\end{equation}
are defined in Eqs.~(\ref{defI32}), (\ref{defb1}), and (\ref{I12}),  respectively.

The series is dominated by a term involving the half-integer power ${(|qH|/T^2)}^{3/2}$, a logarithmic term ${|qH|}^2/T^4 \log |qH|/T^2$ and
two terms quadratic in the magnetic field. If no magnetic field is present, the series reduces to the well-known pion gas contribution
\begin{equation}
P(H=0) = \frac{\pi^2}{30} \, T^4 + {\cal O}(T^8 \log T) \, .
\end{equation}

\section{Conclusions}
\label{conclusions}

Within chiral perturbation theory -- based on a coordinate space representation for the thermal propagators -- we have analyzed the impact
of the magnetic field on the partition function up to the two-loop level. Using the dressed pion masses at zero temperature, we have shown
that the pion-pion interaction in the pressure may be attractive, repulsive, or zero. The respective sign of the two-loop interaction
contribution is controlled by the strength of the magnetic field, as well as temperature and pion mass. In the chiral limit, the
interaction is purely attractive at two-loop order, and gradually becomes weaker as the strength of the magnetic field increases.

We then have provided the low-temperature expansion of the pressure in weak magnetic fields in the chiral limit. The dominant terms in the
series are proportional to ${(|qH|/T^2)}^{3/2}$, $|qH^2|/T^4 \log{|qH|/T^2}$ and $|qH^2|/T^4$ .

The question arises whether three-loop corrections in the thermodynamic quantities -- i.e., order-$p^8$ effects -- are large compared to
the two-loop results discussed here. While the corresponding three-loop analysis referring to zero magnetic field has been provided in
Refs.~\citep{GL89,Hof17}, a three-loop analysis for QCD in presence of a magnetic field, based on chiral perturbation theory, has never been
attempted to the best of our knowledge. Work in this direction, relying on the coordinate space representation, is currently in progress. 

\section*{Acknowledgments}
The author thanks G.\ S.\ Bali, J.\ Bijnens and H.\ Leutwyler for correspondence, as well as R.\ A.\ S\'aenz and C.\ Casta\~no Bernard for
illuminating discussions.

\begin{appendix}

\section{Free Energy Density at Two Loops}
\label{appendixA}

In this appendix we derive the order-$p^6$ contribution to the free energy density, originating from diagrams $6A$-$C$ of
Fig.~\ref{figure1}. The two-loop diagram yields
\begin{equation}
z_{6A} = \frac{M^2}{2 F^2} \, G^{\pm}_1 G^0_1 - \frac{M^2}{8 F^2} \, G^0_1 G^0_1 \, ,
\end{equation}
where the thermal propagators $G^{\pm}_1$ for the charged pions and $G^0_1$ for the neutral pion are defined in Eq.~(\ref{decomposition}).
The result for the one-loop graph 6B,
\begin{equation}
z_{6B} = (4l_5 - 2l_6) \frac{{|qH|}^2}{F^2} \, G^{\pm}_1 + 2 l_3 \frac{M^4}{F^2} \, G^{\pm}_1 + l_3 \frac{M^4}{F^2} \, G^0_1  \, ,
\end{equation}
involves various NLO effective constants $l_i$ that require renormalization (see below). The explicit structure of the tree-level
contribution $z_{6C}$ is not required here, because we are interested in the properties of the system at finite temperature.

In the decomposition of thermal propagators, 
\begin{eqnarray}
& & G^{\pm}_1 = \Delta^{\pm}(0) + {\tilde g}_1(M,T,H) + g_1(M,T,0) \, , \nonumber \\
& & G^0_1 = \Delta^0(0) + g_1(M,T,0) \, ,
\end{eqnarray}
the kinematical functions are finite in the limit $d \to 4$. The zero-temperature propagators $\Delta^{\pm}(0)$ and $\Delta^0(0)$, however,
become singular and take the form
\begin{equation}
\label{DeltaZero}
\Delta^{\pm}(0) = 2 M^2 \lambda + K_1 \, , \qquad \Delta^0(0) = 2 M^2 \lambda\, .
\end{equation}
The integral $K_1$ and the parameter $\lambda$ are
\begin{eqnarray}
\label{defK1}
K_1(M,H) & = & \frac{{|qH|}^{\frac{d}{2}-1}}{{(4 \pi)}^{\frac{d}{2}}} \,  \, {\int}_{\!\!\! 0}^{\infty} \mbox{d} \rho \, \rho^{-\frac{d}{2}+1}
\, \exp\Big( -\frac{M^2}{|qH|} \rho \Big) \, \Big( \frac{1}{\sinh(\rho)} - \frac{1}{\rho} \Big) \, , \nonumber \\
\lambda & = & \mbox{$ \frac{1}{2}$} \, (4 \pi)^{-\frac{d}{2}} \, \Gamma(1-{\mbox{$ \frac{1}{2}$}}d) M^{d-4} \nonumber \\
& & = \frac{M^{d-4}}{16{\pi}^2} \, \Bigg[ \frac{1}{d-4} - \mbox{$ \frac{1}{2}$} \{ \ln{4{\pi}} + {\Gamma}'(1) + 1 \}
+ {\cal O}(d\!-\!4) \Bigg] \, .
\end{eqnarray}

Gathering results, the unrenormalized free energy density at order $p^6$ amounts to
\begin{eqnarray}
\label{z6Unrenormalized}
z^{[6]} & = & z_{6A} +z_{6B} +z_{6C} \nonumber \\
& = &  \frac{3 M^2}{8 F^2} \, {(g_1)}^2 + \frac{M^2}{2 F^2} \, g_1 {\tilde g}_1 \nonumber \\
& & + g_1 \Bigg[ \frac{3 M^4}{2 F^2} \, \lambda + \frac{M^2}{2 F^2} \, K_1 + (4l_5 - 2l_6) \frac{{|qH|}^2}{F^2}
+ 3 l_3 \frac{M^4}{F^2} \Bigg] \nonumber \\
& & + {\tilde g}_1 \Bigg[ \frac{M^4}{F^2} \, \lambda + (4l_5 - 2l_6) \frac{{|qH|}^2}{F^2} + 2 l_3 \frac{M^4}{F^2} \Bigg] \nonumber \\
& & + \frac{3 M^6}{2 F^2} \, \lambda^2 +  \frac{M^4}{F^2} \, K_1 \lambda + (8l_5 - 4l_6) \frac{{|qH|}^2 M^2}{F^2} \, \lambda
+ (4l_5 - 2l_6) \frac{{|qH|}^2}{F^2} \, K_1 \nonumber \\
& & + 6l_3 \frac{M^6}{F^2} \, \lambda + 2 l_3 \frac{M^4}{F^2} \, K_1 + z_{6C} \, .
\end{eqnarray}
The first two terms are quadratic in the kinematical functions and are finite as $d$ approaches the physical dimension $d$=4. Considering
the terms linear in $g_1$ and ${\tilde g}_1$,
\begin{eqnarray}
\label{g1Linear}
& & g_1 \Bigg[ \frac{3 M^4}{2 F^2} \, \lambda + \frac{M^2}{2 F^2} \, K_1 + (4l_5 - 2l_6) \frac{{|qH|}^2}{F^2}
+ 3 l_3 \frac{M^4}{F^2} \Bigg] \, , \nonumber \\
& &  {\tilde g}_1 \Bigg[ \frac{M^4}{F^2} \, \lambda + (4l_5 - 2l_6) \frac{{|qH|}^2}{F^2} + 2 l_3 \frac{M^4}{F^2} \Bigg] \, ,
\end{eqnarray}
using the standard convention for the renormalized NLO effective constants ${\overline l}_i$,
\begin{equation}
\label{NLOCouplings}
l_i = \gamma_i \Big( \lambda + \frac{{\overline l}_i}{32 \pi^2}  \Big) \, , \qquad
\gamma_3 = -\frac{1}{2} \, , \quad \gamma_5 =-\frac{1}{6} \, , \quad \gamma_6 =-\frac{1}{3} \, ,
\end{equation}
we arrive at
\begin{eqnarray}
\label{g1LinearRenorm}
& & + g_1 \Bigg[ -\frac{3 {\overline l}_3}{64 \pi^2} \, \frac{M^4}{F^2} + \frac{M^2}{2 F^2} \, K_1
+ \frac{{\overline l}_6 - {\overline l}_5}{48 \pi^2} \frac{{|qH|}^2}{F^2} \Bigg] \nonumber \\
& & + {\tilde g}_1 \Bigg[ -\frac{{\overline l}_3}{32 \pi^2} \, \frac{M^4}{F^2}
+ \frac{{\overline l}_6 - {\overline l}_5}{48 \pi^2} \frac{{|qH|}^2}{F^2} \Bigg] \, .
\end{eqnarray}
Notice that the above expressions are perfectly finite: all divergences in Eq.~(\ref{g1Linear}) have been canceled. Finally, the
zero-temperature divergences contained in $z_{6A} + z_{6B}$ -- displayed in the last two lines of Eq.~(\ref{z6Unrenormalized}) -- will be
canceled by counterterms from the next-to-next-to-leading order Lagrangian ${\cal L}^6_{eff}$ contained in the zero-temperature
contribution $z_{6C}$.

\section{Kinematical Functions in Weak Magnetic Fields}
\label{appendixB}

In this appendix we provide the low-temperature representations for the kinematical functions in weak magnetic fields. From the very
beginning we operate in the chiral limit. The relevant functions in the free energy density are
\begin{equation}
g_0(0, T, 0) \, , \quad g_1(0, T, 0) \, ,
\end{equation}
that do not involve the magnetic field, and
\begin{equation}
{\tilde g_0}(0, T, H), \quad {\tilde g}_1(0, T, H) \, ,
\end{equation}
that do depend on the magnetic field. The low-temperature analysis for the former functions in the chiral limit has been given a long
time ago in Ref.~\cite{GL89},
\begin{equation}
g_0(0, T, 0) = \frac{\pi^2}{45} \, T^4 \, , \qquad g_1(0, T, 0) = \frac{1}{12} \, T^2\, .
\end{equation}
The latter two functions are defined as
\begin{equation}
\label{defTildegr}
{\tilde g_r}(0, T, H) = \frac{|qH|^{\frac{d}{2} -r}}{{(4 \pi)}^{\frac{d}{2}}} \, {\int}_{\!\!\! 0}^{\infty} \mbox{d} \rho \, \rho^{r-\frac{d}{2}}
\Big( \frac{1}{\sinh(\rho)} - \frac{1}{\rho} \Big) \, \Bigg[ S\Big( \frac{|qH|}{4 \pi T^2 \rho} \Big) -1 \Bigg] \, ,
\end{equation}
with
\begin{equation}
S(z) = \sum_{n=-\infty}^{\infty} \exp(- \pi n^2 z) \, .
\end{equation}
The evaluation of ${\tilde g_1}(0, T, H)$ in weak magnetic fields has been established in Ref.~\citep{Hof19} with the result
\begin{eqnarray}
\label{seriesTildeg1}
{\tilde g_1}(0, T, H) = - \Bigg\{ \frac{|I_{\frac{1}{2}}|}{8 \pi^{3/2}} \sqrt{\epsilon}
- \frac{\log 2}{16 \pi^2} \, \epsilon
+\frac{\zeta(3)}{384 \pi^4} \, \epsilon^2
- \frac{7 \zeta(7)}{98 304 \pi^8} \, \epsilon^4
+ {\cal O}(\epsilon^6) \Bigg\} \, T^2 \, .
\end{eqnarray}
The expansion parameter $\epsilon$ measures the ratio between magnetic field strength and temperature,
\begin{equation}
\epsilon = \frac{|qH|}{T^2} \, .
\end{equation}
By definition, in the weak magnetic field limit $|qH| \ll T^2$, this parameter is small. The integral $I_{\frac{1}{2}}$ is
\begin{equation}
\label{I12}
I_{\frac{1}{2}} = {\int}_{\!\!\! 0}^{\infty} \, \mbox{d} \rho \rho^{-1/2} \Big( \frac{1}{\sinh(\rho)} - \frac{1}{\rho} \Big) \approx -1.516256 \, .
\end{equation}

What remains to be done is the analogous expansion for ${\tilde g_0}(0, T, H)$. According to Ref.~\citep{Hof19}, the representation
(\ref{defTildegr}) can be cast into the form
\begin{eqnarray}
\label{ABC}
{\tilde g_r}(0, T, H) & = & \frac{\epsilon}{{(4 \pi)}^{r+1}} T^{d-2r} \, {\int}_{\!\!\! 0}^1 \mbox{d} \rho \, \rho^{-\frac{d}{2}+r}
\Big( \frac{1}{\sinh(\epsilon \rho/4 \pi)} - \frac{4 \pi}{\epsilon \rho} \Big) \, \Bigg[ S\Big( \frac{1}{\rho} \Big) - 1 \Bigg] \nonumber \\
& & + \frac{\epsilon}{{(4 \pi)}^{r+1}} T^{d-2r} \, \Big\{ I_A + I_B + I_C \Big\} \, ,
\end{eqnarray}
where the respective integrals are defined as
\begin{eqnarray}
\label{IABCexplicit}
I_A & = & {\int}_{\!\!\! 0}^1 \mbox{d} \rho \, \rho^{\frac{d}{2}-r-\frac{5}{2}} \Big( \frac{1}{\sinh(\epsilon/4 \pi \rho)}
- \frac{4 \pi \rho}{\epsilon} \Big)
\Bigg[ S\Big( \frac{1}{\rho} \Big) - 1 \Bigg] \, , \nonumber \\
I_B & = & {\int}_{\!\!\! 0}^1 \mbox{d} \rho \, \rho^{\frac{d}{2}-r-\frac{5}{2}} \Big( \frac{1}{\sinh(\epsilon/4 \pi \rho)}
- \frac{4 \pi \rho}{\epsilon} \Big) \, ,
\nonumber \\
I_C & = & - {\int}_{\!\!\! 0}^1 \mbox{d} \rho \, \rho^{\frac{d}{2}-r-2} \Big( \frac{1}{\sinh(\epsilon /4 \pi \rho)}
- \frac{4 \pi \rho}{\epsilon} \Big) \, .
\end{eqnarray}
For $r$=0 and $d \to 4$, the integral in the first line of Eq.~(\ref{ABC}), much like the integral $I_A$, are well-defined. Following
Ref.~\citep{Hof19}, the integral $I_B$ is split up into two terms,
\begin{eqnarray}
\label{decompositionIB}
I_B & = & I_{B1} + I_{B2} \, , \nonumber \\
I_{B1} & = & \frac{{\epsilon}^{\frac{d}{2}-r-\frac{3}{2}}}{{(4 \pi)}^{\frac{d}{2}-r-\frac{3}{2}}} \, {\int}_{\!\!\! 0}^{\infty} \mbox{d} \rho \,
\rho^{-\frac{d}{2}+r+\frac{1}{2}} \Big( \frac{1}{\sinh(\rho)} - \frac{1}{\rho} \Big) \, , \nonumber \\
I_{B2} & = & - {\int}_{\!\!\! 0}^1 \mbox{d} \rho \, \rho^{-\frac{d}{2}+r+\frac{1}{2}} \Big( \frac{1}{\sinh(\epsilon \rho/4 \pi)}
- \frac{4 \pi}{\epsilon \rho} \Big) \, .
\end{eqnarray}
For $r$=0 and $d \to 4$ we obtain
\begin{eqnarray}
I_{B1} & = & \frac{\sqrt{\epsilon}}{\sqrt{4 \pi}} \, {\int}_{\!\!\! 0}^{\infty} \mbox{d} \rho \,
\rho^{-\frac{3}{2}} \Big( \frac{1}{\sinh(\rho)} - \frac{1}{\rho} \Big) \, , \nonumber \\
I_{B2} & = & - {\int}_{\!\!\! 0}^1 \mbox{d} \rho \, \rho^{-\frac{3}{2}} \Big( \frac{1}{\sinh(\epsilon \rho/4 \pi)}
- \frac{4 \pi}{\epsilon \rho} \Big)  \, .
\end{eqnarray}
Note that the power $\sqrt{\epsilon}$ in $I_{B1}$ is explicit, whereas $\epsilon$ appears in the integrand of $I_{B2}$, as well as in the
integrand in the first line of Eq.~(\ref{ABC}) and in $I_A$ of Eq.~(\ref{IABCexplicit}), as argument of the hyperbolic sine function. We
thus Taylor expand these pieces into
\begin{eqnarray}
\label{taylor}
& & \frac{1}{\sinh(\epsilon \rho/4 \pi)} - \frac{4 \pi}{\epsilon \rho} = c_1 \rho \, \epsilon + c_2 \rho^3 \epsilon^3 + c_3 \rho^5 \epsilon^5
+ {\cal O}(\epsilon^7) \, , \nonumber \\
& & \frac{1}{\sinh(\epsilon/4 \pi \rho)} - \frac{4 \pi \rho}{\epsilon} = c_1 \rho^{-1} \, \epsilon + c_2 \rho^{-3} \epsilon^3
+ c_3 \rho^{-5} \epsilon^5 + {\cal O}(\epsilon^7) \, ,
\end{eqnarray}
such that $\epsilon$-powers in all these integrals become explicit. The first few coefficients $c_p$ in the above series are 
\begin{eqnarray}
& c_1 & = - \frac{1}{24 \pi} \approx -1.33 \times 10^{-2} \, , \nonumber \\
& c_2 & = \frac{7}{23\,040 \, \pi^3} \approx 9.80 \times 10^{-6} \, , \nonumber \\
& c_3 & = - \frac{31}{15\,482\,880 \, \pi^5} \approx -6.54 \times 10^{-9}\, , \nonumber \\
& c_4 & = \frac{127}{9\,909\,043\,200 \, \pi^7}\approx 4.24 \times 10^{-12} \, , \nonumber \\
& c_5 & = -\frac{73}{896\,909\,967\,360 \, \pi^9} \approx -2.73 \times 10^{-15} \, . 
\end{eqnarray}

The last piece in the analysis of ${\tilde g_0}(0, T, H)$ in weak magnetic fields is $I_C$ defined in Eq.~(\ref{IABCexplicit}). This
integral for $r$=0, however, cannot be processed in the manner outlined in Ref.~\citep{Hof19} which indeed worked for the case
$r$=1\footnote{In the decomposition $I_C = I_{C1} +I_{C2}$, Eq.~(A15) of Ref.~\citep{Hof19}, both expressions $I_{C1}$ and $I_{C2}$ are
singular if $r$=0 and $d \to 4$.}. Instead, we decompose the integral $I_C$
\begin{equation}
I_C = - {\int}_{\!\!\! 0}^1 \mbox{d} \rho \Big( \frac{1}{\sinh(\epsilon/4 \pi \rho)} - \frac{4 \pi \rho}{\epsilon} \Big)
\end{equation}
in an alternative way as
\begin{eqnarray}
\label{decompositionIC}
I_C(N) & = & I_{C1}(N) + I_{C2}(N) \\
& = & - {\int}_{\!\!\! 0}^N \mbox{d} \rho \Big( \frac{1}{\sinh(\epsilon/4 \pi \rho)} - \frac{4 \pi \rho}{\epsilon} \Big)
+ {\int}_{\!\!\! 1}^N \mbox{d} \rho \Big( \frac{1}{\sinh(\epsilon/4 \pi \rho)} - \frac{4 \pi \rho}{\epsilon} \Big) \, , \nonumber
\end{eqnarray}
where $N \gg 1$. Redefining integration variables, we obtain
\begin{eqnarray}
I_{C1}(N) & = & - \frac{\epsilon}{4 \pi} \, {\int}_{\!\!\! \epsilon/4 \pi N}^1 \mbox{d} \rho \, \rho^{-2} \, \Big( \frac{1}{\sinh(\rho)}
- \frac{1}{\rho} \Big)
- \frac{\epsilon}{4 \pi} {\int}_{\!\!\! 1}^{\infty} \mbox{d} \rho \rho^{-2} \, \Big( \frac{1}{\sinh(\rho)} - \frac{1}{\rho} \Big) \, ,
\nonumber \\
I_{C2}(N) & = & \frac{\epsilon}{4 \pi} \, {\int}_{\!\!\! \epsilon/4\pi N}^{\epsilon/4\pi} \mbox{d} \rho \, \rho^{-2} \,
\Big( \frac{1}{\sinh(\rho)} - \frac{1}{\rho} \Big) \, .
\end{eqnarray}
The $N$-dependence cancels in the sum $I_{C1}(N) + I_{C2}(N)$, and we are left with
\begin{equation}
I_C = \frac{\epsilon}{4 \pi} \, {\int}_{\!\!\! 1}^{\epsilon/4\pi} \mbox{d} \rho \, \rho^{-2} \, \Big( \frac{1}{\sinh(\rho)} - \frac{1}{\rho}
\Big)
- \frac{\epsilon}{4 \pi} {\int}_{\!\!\! 1}^{\infty} \mbox{d} \rho  \rho^{-2} \, \Big( \frac{1}{\sinh(\rho)} - \frac{1}{\rho} \Big) \, .
\end{equation}
In the second contribution, the power $\epsilon$ is explicit. In the first contribution where $\epsilon$ appears in the upper integration
limit, we Taylor expand the integrand, and then integrate term by term. The final result for $I_C$ can be cast into the form
\begin{equation}
I_C = - \frac{\epsilon}{24 \pi} \, \log\Big(\frac{\epsilon}{4 \pi} \Big)
+ \frac{{\hat J} - {\hat I}}{4 \pi} \, \epsilon
- \sum_{n=2}^{\infty} \frac{2^{2n-1}-1}{(n-1)(2n)!} \, \frac{B_{2n}}{{(4 \pi)}^{2n-1}} \, {\epsilon}^{2n-1} \, ,
\end{equation}
where the $B_{2n}$ are Bernoulli numbers and the quantities $\hat J$ and ${\hat I}$ are defined as
\begin{eqnarray}
{\hat J} & = & \sum_{n=2}^{\infty} \frac{2^{2n-1}-1}{(n-1)(2n)!} \, B_{2n} \approx -0.00924219 \, , \nonumber \\
{\hat I} & = & {\int}_{\!\!\! 1}^{\infty} \mbox{d} \rho \, \rho^{-2} \, \Big( \frac{1}{\sinh(\rho)} - \frac{1}{\rho} \Big) \approx -0.179499
\, .
\end{eqnarray}
Note that the structure of the $\epsilon$-expansion of $I_C$ is now manifest.

Collecting individual contributions, after some algebra, and with the help of the identity
\begin{equation}
\frac{2}{\pi^{\frac{z}{2}}} \, \Gamma\Big(\frac{z}{2}\Big) \zeta(z) = {\int}_{\!\!\! 0}^{\infty} \mbox{d} \rho \, \rho^{\frac{z}{2}-1} \,
\Big[ S(\rho) - 1 \Big] \, ,
\end{equation}
the expansion of the kinematical function ${\tilde g_0}(0, T, H)$ in weak magnetic fields and in the chiral limit takes the form
\begin{eqnarray}
\label{seriesTildeg0}
{\tilde g_0}(0, T, H) & = & \Bigg\{ -\frac{|I_{\frac{3}{2}}|}{8 \pi^{3/2}} \epsilon^{\frac{3}{2}}
- \frac{1}{96 \pi^2} \, \epsilon^2 \, \log\epsilon + b_1 \, \epsilon^2 \nonumber \\
& & + b_2 \, \epsilon^4 + b_3 \, \epsilon^6 + b_4 \, \epsilon^8 + {\cal O}(\epsilon^{10}) \Bigg\} \, T^4 \, ,
\end{eqnarray}
where
\begin{equation}
\label{defI32}
I_{\frac{3}{2}} = {\int}_{\!\!\! 0}^{\infty} \mbox{d} \rho \, \rho^{-\frac{3}{2}} \Big( \frac{1}{\sinh(\rho)} - \frac{1}{\rho} \Big)
\approx -0.610499 \, .
\end{equation}
The coefficient $b_1$ is
\begin{equation}
\label{defb1}
b_1 = \frac{6 ({\hat J} - {\hat I}) - {\tilde I} + \log4\pi}{96 \pi^2} \approx 0.00581159 \, ,
\end{equation}
with
\begin{equation}
{\tilde I} = {\int}_{\!\!\! 0}^1 \mbox{d} \rho \, \Big( \rho^{-1} + \rho^{-\frac{3}{2}} \Big) \,  \Bigg[ S\Big( \frac{1}{\rho} \Big) - 1 \Bigg]
- {\int}_{\!\!\! 0}^1 \mbox{d} \rho \, \rho^{-\frac{1}{2}} \, ,
\end{equation}
while the coefficients $b_p \, (p \ge 2)$ are
\begin{equation}
\label{definitionBp}
b_p = -\frac{2 (2^{2p-1}-1)}{{(4\pi)}^{2p} (2p)!} \,
\Bigg\{ \frac{2 \Gamma( 2p-\mbox{$\frac{3}{2}$}) \zeta(4p -3)}{\pi^{2p-\frac{3}{2}}} + \frac{1}{1-p} \Bigg\}  \, B_{2p} \, ,
\quad p \ge 2 \, .
\end{equation}
The numerical values of the first five coefficients $b_p \, (p \ge 2)$ are given in Table \ref{table1}.

\begin{table}[ht!]
\centering
\begin{tabular}{|c|c|}
\hline
$p$  &  $b_p$ \\
\hline
2  &  - 6.56867042287 $\times 10^{-7}$  \\
\hline
3  &    1.90033315207 $\times 10^{-10}$  \\
\hline
4  &    1.55270844266 $\times 10^{-15}$  \\
\hline
5  & -  3.08314759762 $\times 10^{-16}$  \\
\hline
6  &    1.87712447343 $\times 10^{-18}$  \\
\hline
\end{tabular}
\caption{The first five coefficients $b_p$ defined by Eq.~(\ref{definitionBp}).}
\label{table1}
\end{table}

More explicitly, the series can be written as
\begin{eqnarray}
\label{seriesTildeg0}
{\tilde g_0}(0, T, H) & = & \Bigg\{ -\frac{|I_{\frac{3}{2}}|}{8 \pi^{3/2}} \epsilon^{\frac{3}{2}}
- \frac{1}{96 \pi^2} \, \epsilon^2 \, \log\epsilon
+ \frac{6 ({\hat J} - {\hat I}) - {\tilde I} + \log4\pi}{96 \pi^2} \, \epsilon^2 \nonumber \\
& & - \frac{7 (2 \pi^2 - 3 \zeta(5))}{184320 \pi^6} \, \epsilon^4
+ \frac{31 (4 \pi^4 - 105 \zeta(9))}{495452160 \pi^{10}} \, \epsilon^6 \nonumber \\
& & - \frac{127 (32 \pi^6 - 31185 \zeta(13))}{3805072588800 \pi^{14}} \, \epsilon^8 + {\cal O}(\epsilon^{10}) \Bigg\} \, T^4 \, .
\end{eqnarray}

\begin{table}[ht!]
\centering
\begin{tabular}{|c||c|c|c|c|}
\hline
$\epsilon$ & ${\tilde g_0}/T^4$ &  ${\cal O}({\epsilon}^{3/2})$ &  ${\cal O}(\epsilon^2 \log\epsilon)$ &  ${\cal O}(\epsilon^2)$ \\
\hline
\hline
0.1    & -3.50963233726 -4  & -4.33381291264 -4  & -4.09079140529 -4  & -3.50963246014 -4 \\
\hline
0.05   & -1.30789993783 -4  & -1.53223424946 -4  & -1.45318968180 -4  & -1.30789994551 -4 \\
\hline
0.01   & -1.26375177959 -5  & -1.37047197570 -5  & -1.32186767423 -5  & -1.26375177971 -5 \\
\hline
0.005  & -4.56026045635 -6  & -4.84535013722 -6  & -4.70555019271 -6  & -4.56026045643 -6 \\
\hline
0.001  & -4.20279056592 -7  & -4.33381291264 -7  & -4.26090646044 -7  & -4.20279056592 -7 \\
\hline 
0.0005 & -1.49764974370 -7  & -1.53223424946 -7  & -1.51217871733 -7  & -1.49764974370 -7 \\
\hline
0.0001 & -1.35493952595 -8  & -1.37047197570 -8  & -1.36075111541 -8  & -1.35493952595 -8 \\
\hline
\end{tabular}
\caption{Exact result and leading terms in the series (\ref{seriesTildeg0}) for the kinematical function ${\tilde g_0}$ in the limit
$|qH| \ll T^2$. We use the notation where $-3.50963233726 -4$ stands for $-3.50963233726 \times 10^{-4}$, etc.}
\label{table2}
\end{table}

To check convergence properties of the above series for ${\tilde g_0}(0,T,H)$ in the weak magnetic field limit $|qH| \ll T^2$, let us
compare the first few terms in the $\epsilon$-expansion with the exact result Eq.~(\ref{defTildegr}). The first column in Table
\ref{table2} displays the exact result, while the second column just takes into account the leading term in the series
(\ref{seriesTildeg0}) proportional to ${\epsilon}^{3/2}$. The third column furthermore incorporates the
$\epsilon^2 \log\epsilon$-contribution and the fourth column finally extends up to the  $\epsilon^2$-term. One observes that a very good
approximation is achieved by just including the first three terms: the series (\ref{seriesTildeg0}) converges quite rapidly.

Finally it should be noted that the order $T^4$-contribution in the pressure, i.e., the coefficient $p_1$ in Eq.~(\ref{coeffp1p2}),
\begin{equation}
p_1(t,m,m_H) = h_0(M^{\pm}_{\pi},T,0) + \mbox{$ \frac{1}{2}$} h_0(M^0_{\pi},T,0) + {\tilde h}_0(M^{\pm}_{\pi},T,H) \, ,
\end{equation}
contains the kinematical functions $h_0(M^{\pm}_{\pi},T,0)$ and ${\tilde h}_0(M^{\pm}_{\pi},T,H)$ which, in the chiral limit, reduce to
\begin{equation}
h_0(M_H,T,0) \, , \qquad  {\tilde h}_0(M_H,T,H) \, ,
\end{equation}
with
\begin{equation}
{(M_H)}^2 = \frac{{\overline l_6} - {\overline l_5}}{48 \pi^2} \, \frac{q^2 H^2}{F^2} \, .
\end{equation}
In the weak magnetic field limit, the kinematical function $h_0(M_H,T,0)$ hence takes the form
\begin{equation}
h_0(M_H,T,0) = h_0(0,T,0) - \alpha \epsilon^2 h_1(0,T,0) + \frac{\alpha^2 \epsilon^4}{2} \, h_2(0,T,0) + {\cal O}(\epsilon^6) \, ,
\end{equation}
where
\begin{equation}
\alpha = \frac{{\overline l_6} - {\overline l_5}}{12 \pi} \, t^2 \, , \qquad \epsilon = \frac{|qH|}{T^2} \, ,
\qquad t = \frac{T}{4 \pi F} \, .
\end{equation}
Analogously, in the weak magnetic field limit, the kinematical function ${\tilde h}_0(M_H,T,H)$ amounts to
\begin{equation}
{\tilde h}_0(M_H,T,H) = {\tilde h}_0(0,T,H) - \alpha \epsilon^2 {\tilde h}_1(0,T,H) + \frac{\alpha^2 \epsilon^4}{2} \, {\tilde h}_2(0,T,H)
+ {\cal O}(\epsilon^6) \, .
\end{equation}
We hence have additional terms in the weak magnetic field expansion of the pressure in the chiral limit, which contain the NLO low-energy
constants $\overline l_5$ and $\overline l_6$.

\end{appendix}

\end{document}